\documentclass[reprint,amsmath,amssymb,aps,prl]{revtex4-1}
\usepackage{graphicx}
\usepackage{dcolumn}
\usepackage{bm}
\usepackage[colorlinks]{hyperref}
\usepackage[caption=false]{subfig}
 
\graphicspath{{Figures/}}
\usepackage[dvipsnames]{xcolor}
\usepackage[final]{changes}
\definechangesauthor[name=Michael Wiedmann,color=blue]{MW}
\definechangesauthor[name=Joachim Ankerhold,color=PineGreen]{JA}
\definechangesauthor[name=Juergen Stockburger,color=magenta]{JS}

\begin{document}
\sloppy

\title{Out-of-equilibrium operation of a quantum heat engine:\\ The cost of thermal coupling control}

\author{Michael Wiedmann}
\author{J\"urgen T. Stockburger}
\author{Joachim Ankerhold}
\affiliation{Institute for Complex Quantum Systems and IQST, University of Ulm, 89069 Ulm, Germany}

\date{\today}

\begin{abstract}
Real quantum heat engines lack the separation of time and length scales that is characteristic for classical engines. They must be understood as open quantum systems in non-equilibrium with time-controlled coupling to thermal reservoirs as integral part. Here, we present a systematic approach to describe a broad class of engines and protocols beyond conventional weak coupling treatments
starting from a microscopic modeling. For the four stroke Otto engine the full dynamical range down to low temperatures is explored and the crucial role of the work associated with the coupling/de-coupling to/from reservoirs in the energy balance is revealed. Quantum correlations turn out to be instrumental to enhance the efficiency which opens new ways for optimal control techniques. 
\end{abstract}

\maketitle

\emph{Introduction}--Macroscopic thermodynamics was developed for very practical reasons, namely, to understand and describe the fundamental limits of converting heat into useful work. In \emph{ideal} heat engines, components are always in perfect thermal contact or perfectly insulated, resulting in reversible operation. The work medium of real macroscopic engines is typically between these limits, but internally equilibrated, providing finite power at a reduced efficiency. Any reduction of engine size to microscopic dimensions calls even this assumption into doubt.

At atomic scales and low temperatures, quantum mechanics takes over, and concepts of classical thermodynamics may need to be modified \cite{mahler2009,mukamel2009,hanggi2011,jennings17,eisert18}. This is not only of pure theoretical interest but has immediate consequences in the context of recent progress in fabricating and controlling thermal quantum devices \cite{nori2007,popescu2010,kurizki2013,pekola2015}. While the first heat engines implemented with trapped ions \cite{abah12,rossn16} or solid state circuits \cite{pekola2014} still operated in the classical regime, more recent experiments entered the quantum domain~\cite{mottonen2017,pekola2018,klatz19,schmidtkaler2018}. In the extreme limit, the work medium may even consist of only a single quantum object \cite{PhysRevX.5.031044}.

\begin{figure}[t]
	\centering
		\includegraphics[width =0.775\columnwidth]{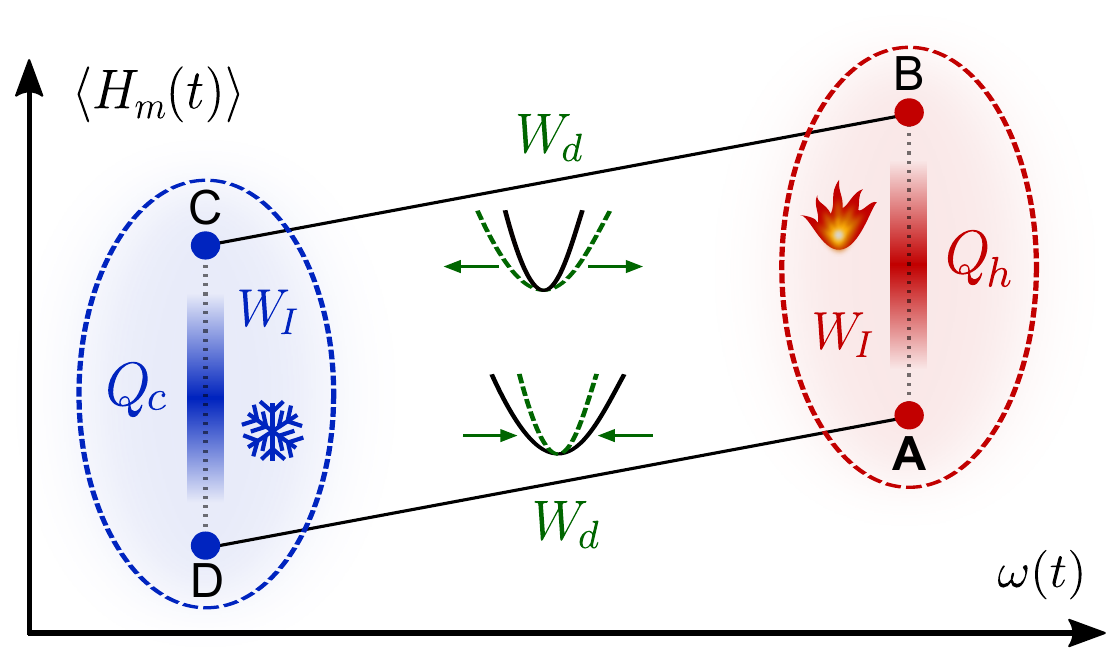}
		\break
	\centering
	    \includegraphics[width =0.8\columnwidth]{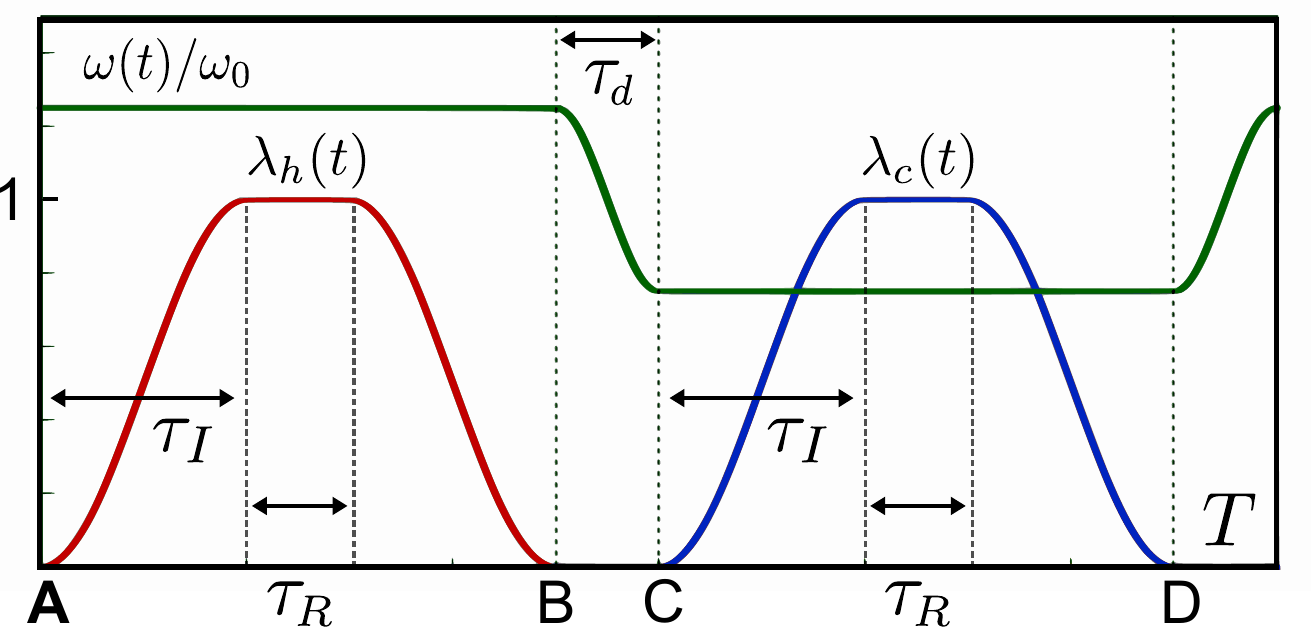}
			\caption{\label{fig2}   Top: Energy-frequency diagram of the work medium in a quantum Otto heat engine with frequency $\omega(t)$ varying around $\omega_0$. A cycle includes two isochore ($A\to B, C\to D$) and two isentropic strokes ($B\to C, D\to A$). Bottom: Thermal contact to hot (cold) reservoirs is controlled by $\lambda_h(t)$ $[\lambda_c(t)]$ and expansion (compression) is due to $\omega(t)$. The cycle is specified by three characteristic time scales $\tau_I$, $\tau_d$, $\tau_{R}$.}
			\label{fig1}
\end{figure}

Theoretically, one is thus faced with the fundamental challenge that a separation of time and length scales on which conventional descriptions of thermal engines is based, may no longer apply. This has crucial consequences: First, the engine's operation must be understood as a specific mode of the cyclic \emph{dynamics} of an open quantum system with the coupling/de-coupling processes to/from thermal reservoirs being integral parts of the time evolution; second, thermal coupling strength and thermal times $\hbar/k_{\rm B} T$ at low temperatures $T$ may match characteristic scales of the work medium.  The latter requires a non-perturbative treatment beyond standard weak-coupling approaches \cite{gonza18,camat18,alicki1979,geva1992,Rezek_2006,kosloff2017,hekking2013,horowitz2013,koslo13,esposito2015,PhysRevX.5.031044,PhysRevA.97.062121,PhysRevA.97.032104,hofer16,roule18a} to include medium-reservoir quantum correlations and non-Markovian effects \cite{newma17,PhysRevE.98.032121,ankerhold2014,eisert2014,bera2017,Pezzutto_2019,abah19a}. The former implies that  coupling work $W_I$ may turn into an essential ingredient in the energy balance whereas typically only exchanged heat $Q$ and work for compression/expansion of the medium (driving work) $W_d$ is adressed, see Fig.~\ref{fig2}. Roughly speaking, while in a classical engine the cylinder is much bigger than the valve so that $|W_d|\gg |W_I|$, for a quantum device this separation of scales may fail and $|W_d|\sim |W_I|$. This is particularly true for scenarios to approach the quantum speed limit in cyclic operation \cite{deffner2017,abah2019b,campo14}. Can a quantum engine under these conditions be operated at all?

To adress this question, in this Letter we push forward a {\em non-perturbative} treatment and apply it to a finite-time generalization of the Otto cycle (Fig.~\ref{fig2}). It is based on an exact mapping of the Feynman-Vernon path integral formulation \cite{feynm63,weiss12} onto a Stochastic Liouville-von Neumann equation \cite{stock02} which has been successfully applied before \cite{stock99a,schmi11,schmi13,schmi15,Motz_2018}. Here, we extend it to accommodate time-controlled thermal contact between medium and reservoirs and thus to arrive at a systematic treatment of quantum heat engines at low temperatures, stronger coupling and driving. Work media with either a single harmonic or anharmonic degree of freedom are discussed to make contact with current experiments. We demonstrate the decisive role of the {\em coupling work} $W_I$ which inevitably must enter the energy balance. Its dependence on quantum correlations opens ways for optimal control \cite{schmi11}.

\emph{Modeling}--A quantum thermodynamic device with cyclic operation involving external work and two thermal reservoirs is described by the generic Hamiltonian
\begin{equation*}
    H(t)=H_m(t)+  H_{c} + H_{I, {c}}(t) +  H_{h} + H_{I, {h}}(t)\,,
\end{equation*}
where $H_m, H_{c/h}$ denote the Hamiltonians of the work medium and the cold/hot reservoirs, respectively, with interactions $H_{I, c/h}$. Not only the working medium is subject to external control, but also the couplings---this is required
in a full dynamical description of the compound according to specific engine protocols. We consider a particle in a one-dimensional potential, $H_m(t) = p^2/(2 m) + V(q,t)$, also motivated by recent ion-trap experiments~\cite{abah12,rossn16}. Reservoirs are characterized not only by their temperatures $T_c<T_h$, but also by a coupling-weighted spectral density~\cite{calde83,weiss12}. Assuming the free fluctuations of each reservoir to be Gaussian, these can be modeled in a standard way \cite{weiss12} as a large collection of independent effective bosonic modes with coupling terms of the form
$H_{I, c/h}(t)=-\lambda_{c/h}(t) q\sum_k c_{k, c/h} (b_{k, c/h}^\dagger+b_{k,c/h})+\frac{1}{2} q^2 \lambda_{c/h}^2 (t) \mu_{c/h}$. The coefficients $\mu_{c/h}$ are conventionally chosen such that only the dynamical impact of the medium-reservoir coupling matters \cite{calde83}. In a quasi-continuum limit the reservoirs become  infinite in size; thermal initial conditions are therefore sufficient to ascertain their roles as heat baths.

The dynamics of this setting will be explored over sufficiently long times such that a regime of periodic operation is reached, without limitations on the ranges of temperature, driving frequency, and system-reservoir coupling strength. The nature of the quantum states encountered, either as quantum heat engine (QH) or refrigerator (QR), is not known a priori. 

In order to tackle this formidable task, we start from the Feynman-Vernon path integral formulation \cite{feynm63,calde83,weiss12}. It provides a formally exact expression for the reduced density operator $\rho_m(t)={\rm Tr}_R\{\rho_{\rm tot}(t)\}$ of the working medium. The quantum correlation functions $L_{c/h}(t-t')=\langle X_{c/h}(t) X_{c/h}(t')\rangle$ with $X_{c/h} = \sum_k c_{k, c/h} (b_{k, c/h}^\dagger+b_{k,c/h})$ are memory kernels of a \emph{non-local} action functional, representing the influence of the reservoir dynamics on the distinguished system as a retarded self-interaction. This formulation can be exactly mapped onto a Stochastic Liouville-von Neumann equation (SLN) 
\cite{stock02}, an approach which remains consistent in the regimes of strong coupling, fast driving, and low temperatures \cite{stock99a,schmi11,schmi13, schmi15,Motz_2018}, where master equations become speculative or inaccurate.

Here we extend an SLN-type method for ohmic dissipation~\cite{stock99}
to time-dependent control of the system-reservoir couplings, including stochastic representations of key reservoir observables \cite{SM}. The resulting dynamics is given by
\begin{equation}
\dot{\rho}_\xi(t)=-\frac{i}{\hbar}[H_m(t),\rho_\xi]+\mathcal{L}_h[\rho_\xi]+\mathcal{L}_c[\rho_\xi]
\label{SLED}
\end{equation}
which contains, in addition to terms known from the master equation of Caldeira and Leggett~\cite{calde83b}, further terms related to the control of system-reservoir couplings and to finite-memory quantum noise $\xi_{\rm c/h}$,
\begin{eqnarray}
    \mathcal{L}_{\alpha} &=&  - \frac{m\gamma_{\alpha}}{2\hbar^2} \lambda_{\alpha}^2(t)\big[q, i\{p,\rho_{\xi}\}+ 2 k_{\rm B} T_{\alpha} [q,\rho_{\xi}]\big]\nonumber\\
&&- \frac{i}{\hbar}\lambda_{\alpha}(t)\left\{ \frac{m \gamma_{\alpha}}{2\hbar} \dot{\lambda}_{\alpha}(t)[q^2,\rho_{\xi}]- \xi_{\alpha}(t) [q,\rho_{\xi}]\right\}\, .
\label{dissipators}
\end{eqnarray}
Averaging over samples of the operator-valued process $\rho_\xi(t)$ yields the physical reduced density $\rho_m(t)=\mathbb{E}[\rho_\xi(t)]$. The independent noise sources $\xi_\alpha(t)$ are related to the reservoir correlation functions through $\langle \xi_\alpha(t)\xi_\alpha(t')\rangle = \Re L_\alpha(t-t')-\frac{2m\gamma_\alpha}{\hbar\beta_\alpha}\delta(t-t')$.

The time local Eqs. (\ref{SLED}) and (\ref{dissipators}) thus provide a {\em non-perturbative,  non-Markovian} simulation platform  for quantum engines with working media consisting of single or few continuous or discrete (spin) degrees of freedom; different protocols can be applied with unambiguous identification of per-cycle energy transfers to work or heat reservoirs. Next, we will apply it to a four stroke Otto cycle.

\emph{Engine cycle}--For this purpose, steering of both the time-dependent potential $V(q, t)$ and time-dependent couplings $\lambda_{c/h}(t)$ in an alternate mode is implemented, see Fig.~\ref{fig1}. For simplicity, ohmic reservoirs with equal damping rate $\gamma$ are assumed. A single oscillator degree of freedom represents the working medium as a particle moving in
\begin{equation}
    V(q,t)=\frac{1}{2} m \omega^2(t) q^2 + \frac{1}{4} m \kappa q^4
    \label{eq:pot}
\end{equation}
with a parametric-type of driving $\omega(t)$ and anharmonicity parameter $\kappa\geq 0$. We consider $\omega(t)$ varying around a center frequency $\omega_0$  between $\omega_0\pm \frac{\Delta\omega}{2}, (\Delta\omega>0)$, within the time $\tau_d$ during the isentropic strokes of expansion (B $\to $ C) and compression (D $\to $ A); it is kept constant along the hot and cold isochores (A $\to$ B and C $\to $ D) (cf.\ Fig.~\ref{fig1}). The isochore strokes are divided into an initial phase raising the coupling parameter $\lambda_{c/h}$ from zero to one with duration $\tau_{I}$, a relaxation phase of duration $\tau_R$, and a final phase with $\lambda_{c/h}\to 0$, also of duration $\tau_I$.

The cycle period is thus $T = 4\tau_I + 2\tau_d + 2\tau_{R}$, as indicated in Fig.~\ref{fig1}. The total simulation time covers a sufficiently large number of cycles to approach a periodic steady state (PSS)  with $\rho_m(t)=\rho_m(t+T)$. Conventionally, one neglects what happens during $\tau_I$;  one assumes that modulating the thermal interaction has no effect on the energy balance (see also Ref.~\cite{newma17}). In the quantum regime, such effects may, on the contrary, play a crucial role as will be revealed in the sequel. 

\begin{figure}[b]
	\centering
 		\includegraphics[width=\columnwidth]{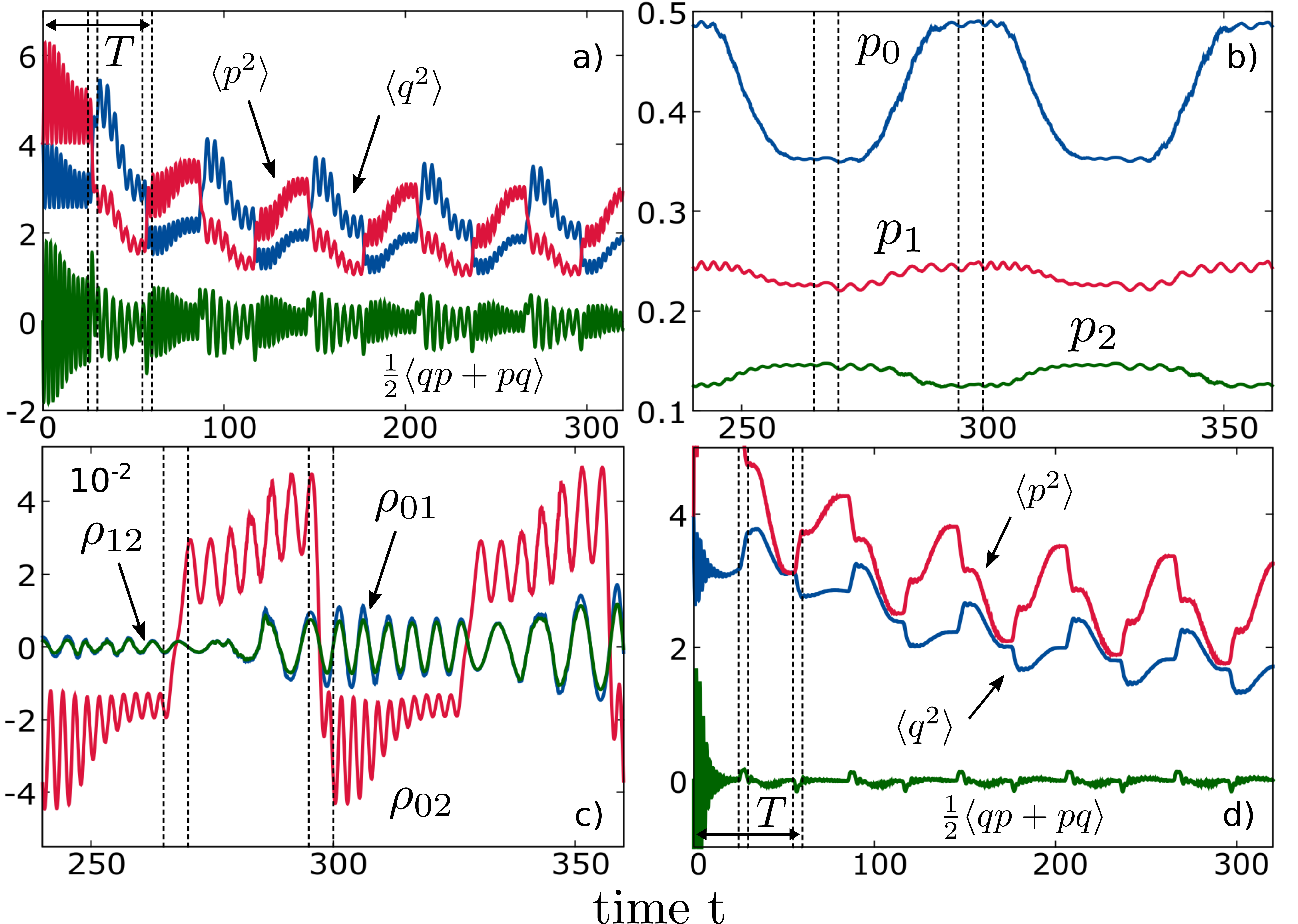}
		\caption{\label{fig:populations} Quantum dynamics for an Otto engine with $\omega_0\hbar\beta_h = 0.25$, $\omega_0\hbar\beta_c = 3$. Time scales are $\omega_0\tau_I = 10$, $\omega_0\tau_{d} = 5$, $\omega_0 T = 60$ with reservoir coupling $\gamma/\omega_0 = 0.05$; here and in the sequel $\omega_{\rm cut}/\omega_0 = 30$.
		(a) Harmonic and (b) anharmonic ($\kappa = 0.15$) work medium: Approach of a PSS for the variances in position, momentum and the cross-correlations $\langle qp + pq\rangle$. (c) Harmonic Fock state populations $p_j(t)$ at frequency $\omega_0$ and (d) off-diagonal elements $\Re\rho_{ij}(t)$ (coherences), see text for details.}
\end{figure}

\emph{Periodic steady state}--In Fig.~\ref{fig:populations}(a-c) results are shown for a purely harmonic system, for which analytical results have been derived in limiting cases~\cite{abah12,abah_2016,PhysRevE.98.032121}. We use it as a starting point to refer to the situation in ion trap experiments \cite{rossn16} and to identify in (d) the role of anharmonicities. After an interval of transient dynamics (a), the elements of the covariance matrix  settle into a time-periodic pattern with damped oscillations near frequencies $\omega_0\pm\frac{\Delta\omega}{2}$; the  time to reach a PSS typically exceeds a single period. The presence of $qp$-correlations manifests broken time-reversibility which implies that a description in terms of stationary distributions with effective temperatures is not possible.  Indeed, the PSS substantially deviates from a mere sequence of equilibrium states as also illustrated by the von Neumann entropy $S_{vN}$ \cite{SM}.
Further insight is gained by taking the oscillator at its mean frequency $\omega_0$ as a reference and employ the corresponding Fock state basis to monitor populations $p_n(t)=\langle n|\rho(t)|n\rangle$ and coherences $\rho_{nm}(t)=\langle n|\rho(t)|m\rangle, n\neq m$ (b, c). Population from the ground and the first excited state is transferred to (from) higher lying ones during contact with the hot (cold) bath. In parallel, off-diagonal elements $\rho_{nm}(t)$ are maintained. These are dominated by $\rho_{02}(t)$ contributions according to the parametric-type of driving during the isentropic strokes. While this Fock state picture has to be taken with some care for dissipative systems, it clearly indicates the presence of coherences associated with $qp$-correlations in the medium \cite{koslo13,anders2016}. The impact of anharmonicities for stiffer potentials in (\ref{eq:pot}) is depicted in (d). In comparison to the harmonic case, dynamical features display smoother traces with enhanced (reduced) variations in $\langle p^2\rangle$ ($\langle q^2\rangle$). Non-equidistant energy level spacings may in turn influence the efficiency (see below).

\emph{Work and heat}--The key thermodynamic quantities of a QH are work and heat per cycle. Note that even though we operate the model with a medium far from equilibrium, these quantities have a sound and unique definition in the context of fully Hamiltonian dynamics involving reservoirs of infinite size. An assignment of separate contributions of each stroke to heat and work is not needed in this context. Moreover, any such assignment in a system with finite coupling would raise difficult conceptual questions due to system-reservoir correlations \cite{talkn16,bera2017}.

\replaced[id=JS]{In the context of full system-reservoir dynamics, heat per steady-state cycle is uniquely defined as the energy change of the reservoir}{During the isochores heat is exchanged between medium and reservoirs with the heat }
\begin{equation*}
Q_{c/h} = - \int_0^T dt\,  {\rm Tr}\{H_{c/h}\,  \dot{\rho}_{\rm tot}(t)\}\, 
\added[id=JS]{.}
\end{equation*}
\deleted[id=JS]{released from a reservoir per cycle.}
Within the SLN the integrand consists of separate terms associated with reservoir noise and medium back action~\cite{SM}.

\replaced[id=JS]{Similarly, work}{Work} is \replaced[id=JS]{obtained}{measured} as injected power, i.e., 
\begin{equation*}
    W=\int_0^T dt\, {\rm Tr}\left\{ \frac{\partial H(t)}{\partial t}\rho_{\rm tot}(t)\right\} \, ,
\end{equation*}
\replaced[id=JS]{where separate driving and coupling work contributions $W_d$ and $W_I$ can be discerned from $\partial_t H = \partial_t H_m + \partial_t H_{I,c} + \partial_t H_{I,h}$.}
{\begin{equation*}
    W_d=\int_0^T dt\, {\rm Tr}\left\{ \frac{\partial H_m(t)}{\partial t}\rho_{\rm tot}(t)\right\} \, ,
\end{equation*}
while work associated with the coupling/de-coupling processes between medium and respective baths $W_{I, {c/h}}$ (coupling work) during the isochores is obtained in the same way from $H_{I, {c/h}}$. Potential ambiguities in thermodynamic quantities arising from finite coupling energies $\langle H_{I, {c/h}}\rangle$~\cite{koslo13,seife16,talkn16,bera2017} do not arise in a complete PSS cycle.}\deleted[id=JS]{While ambiguities in the definition of the internal energies of system and reservoirs can arise from finite terms $\langle H_{I, {c/h}}\rangle $~\cite{koslo13,seife16,talkn16,bera2017}, these terms do not enter heat and work of a complete cycle.} Careful analysis indicates \cite{SM} that the coupling work $W_I$ is completely dissipated~\footnote{This has been previously shown for a Markovian classical heat engine model~\cite{Aurel17}}. 

\begin{figure}[t]
	\centering
 		\includegraphics[width=\columnwidth]{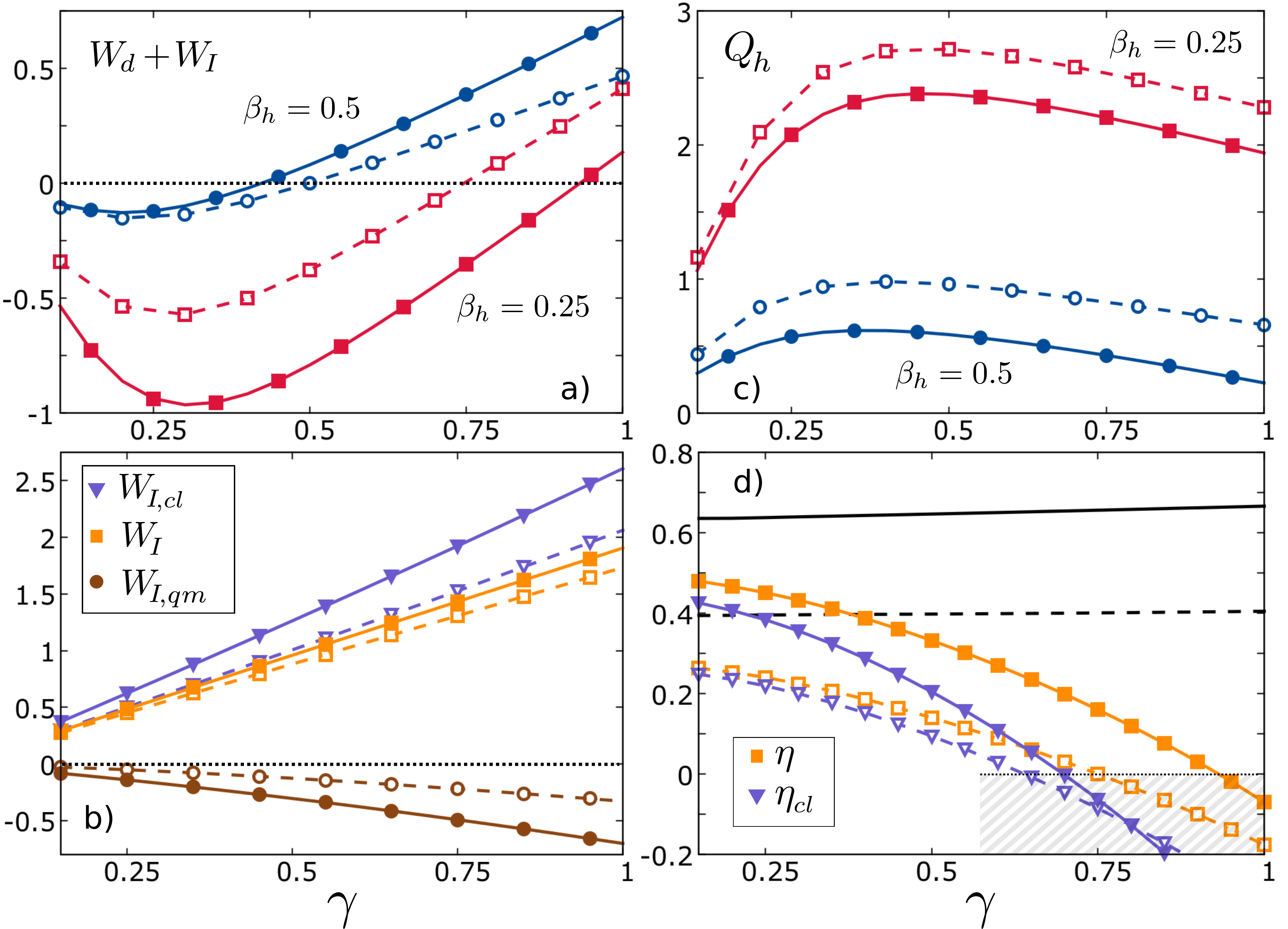}
		\caption{\label{fig:fullcycle} Thermodynamic quantities of the quantum Otto cycle vs.\ thermal coupling $\gamma/\omega_0$ at $\omega_0\hbar\beta_c = 3$. Process times are  $\omega_0\tau_I = \omega_0\tau_{d} = 5$,  $\omega_0 T = 40$ with compression $\Delta \omega/\omega_0 = 1$ (solid),  0.5 (dashed). (a) Net work $W_d+W_I$ and (b) contributions of coupling work $W_I=W_{I, cl}+ W_{I, qm}$; (c) absorbed heat $Q_h$;  (d) efficiency $\eta$ according to (\ref{HEeff}) (orange), ignoring $W_I$ (black), and with only $W_{I, cl}$ included (violet); the dissipator phase is the dashed area. (b,d) at fixed $\hbar\omega_0\beta_h=0.25$}
\end{figure}

Figure~\ref{fig:fullcycle} (a) displays the strong coupling dependence of the net work $W_d+W_I$. It turns from negative (net work output) to positive, thus highlighting the coupling work $W_I$ as an essential contribution in the work balance. The SLN approach allows to reveal its two parts, i.e. $W_I=W_{I, cl}+W_{I, qm}$, where the first one, determined by $\langle q^2\rangle$, also exists at high temperatures while the second one is a genuine quantum part depending on $qp$-correlations. One can show that $W_{I, cl}>0$ dominates  while, with increasing compression rate $\frac{\Delta\omega}{\omega_0}$, $W_{I, qm}$ contributes substantially with a sign depending on the {\em phase} of the $qp$-correlations\added[id=JS]{ relative to the timing of the coupling control}. By choosing $\tau_d,\tau_I$ as in Fig.~\ref{fig:fullcycle}, one achieves $W_{I, qm}<0$, thus counteracting $W_{I, cl}$ (b). In turn, control of $qp$-quantum correlations opens ways to tune the impact of $W_I$ on the energy balance \cite{SM}. Heat $Q_{h}$, see (c),  follows a non-monotonous behavior with $\gamma$, also a genuine quantum effect that cannot be captured by standard weak coupling approaches. Its decrease beyond a maximum can be traced back to enhanced momentum fluctuations due to damping.

We are now in a position to discuss the ratio
\begin{equation}
    \eta = - \frac{W_d+W_I}{Q_{\rm h}}\,
    \label{HEeff}
\end{equation}
which describes the efficiency of a QH if $W_d+W_I<0$\deleted[id=JS]{ since $Q_h>0$}. 

\begin{figure}[t]
	\centering
        \includegraphics[width =0.75\columnwidth]{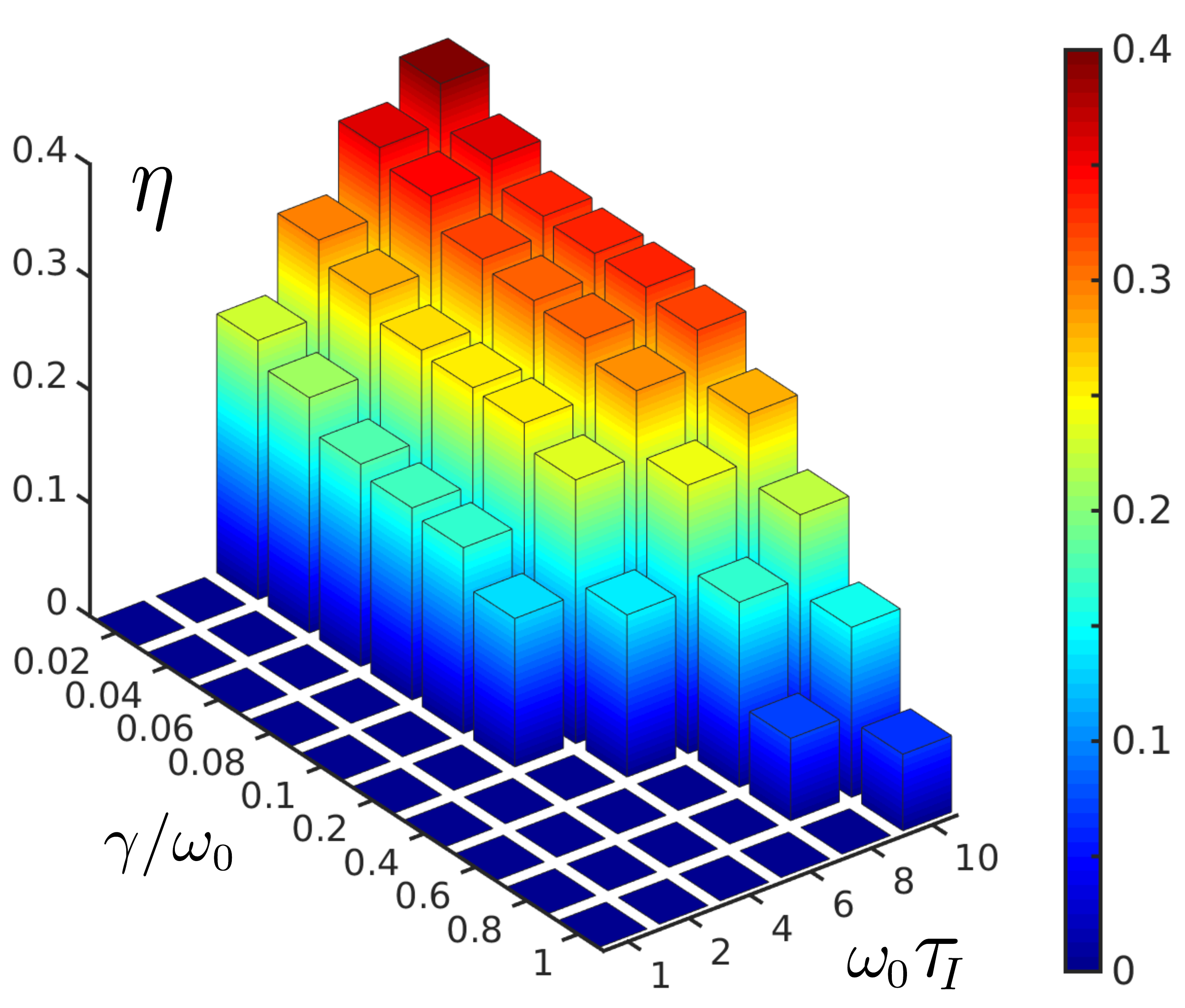}
        \caption{\label{fig:phasediagram} The PSS's operating phases as a QH ($\eta>0$) and as a dissipator ($\eta$ set to 0) vs.\ medium-reservoir coupling and adiabaticity parameter. Other parameters are $\omega_0\hbar\beta_h = 0.5 $, $\omega_0\hbar\beta_c = 3$, $\omega_0\tau_{d}=\omega_0\tau_R=5$.}
\end{figure}

In regimes where $\eta$ is nominally negative, the system is not a QH, but merely a dissipator. The theory of the adiabatic Otto cycle and its extension using an adiabaticity parameter predicts some regimes of pure dissipation~\footnote{See Eq. (6) of \cite{abah12}.}, however, without recognizing the coupling work $W_I$ as an essential ingredient. As seen above, its detrimental impact can be soothed by quantum correlations, see (d). 

The combined dependence on $\gamma$ and thermalization adiabaticity parameter $\omega_0\tau_I$ yields a phase diagram pointing out QH phase ($\eta>0$) and a dissipator phase ($\eta \leq 0$) over a broad range of thermal couplings up to $\gamma/\omega_0\sim 1$, Fig.~\ref{fig:phasediagram}. A QH is only realized if $\tau_I$ exceeds a certain threshold which \emph{grows} with increasing medium-reservoir coupling. To make this more quantitative, progress is achieved for small compression ratios to estimate $W_d$ and $W_I$ and derive \cite{SM} from $W_d+W_I<0$ the relation
\begin{equation}
\omega_0\tau_I > \frac{2\gamma}{\Delta\omega} \frac{1+R}{1-R}\, .
\end{equation}
\replaced[id=JS]{with $R =(\langle q^2\rangle_A+\langle q^2\rangle_D)/(\langle q^2\rangle_C+\langle q^2\rangle_B)$}{Here, $0\leq R\leq 1$ is determined by the variances in position at the cycle times A--D in} and where $0\leq R \leq 1$. Since $W_d\sim \dot{\omega}(t) \tau_d\sim \Delta\omega$ and $W_I\sim \dot{\lambda}^2(t)\tau_I\sim 1/\tau_I$, for short cycle times $W_I$ always dominates.
Qualitatively, for the parameters in Fig.~\ref{fig:phasediagram}, very weak coupling $\gamma\tau_R,\gamma\tau_I\ll 1$ leads to $(1-R)/\gamma\approx const.$ and thus $\omega_0\tau_I\approx const.$; for larger coupling with more efficient heat exchange, the $\gamma$ dependence of $R$ is less relevant \cite{SM} so that $\omega_0\tau_I\sim\gamma/\Delta\omega$ as in Fig.~\ref{fig:phasediagram}. 

\begin{figure}[b]
	\centering
 		\includegraphics[width=\columnwidth]{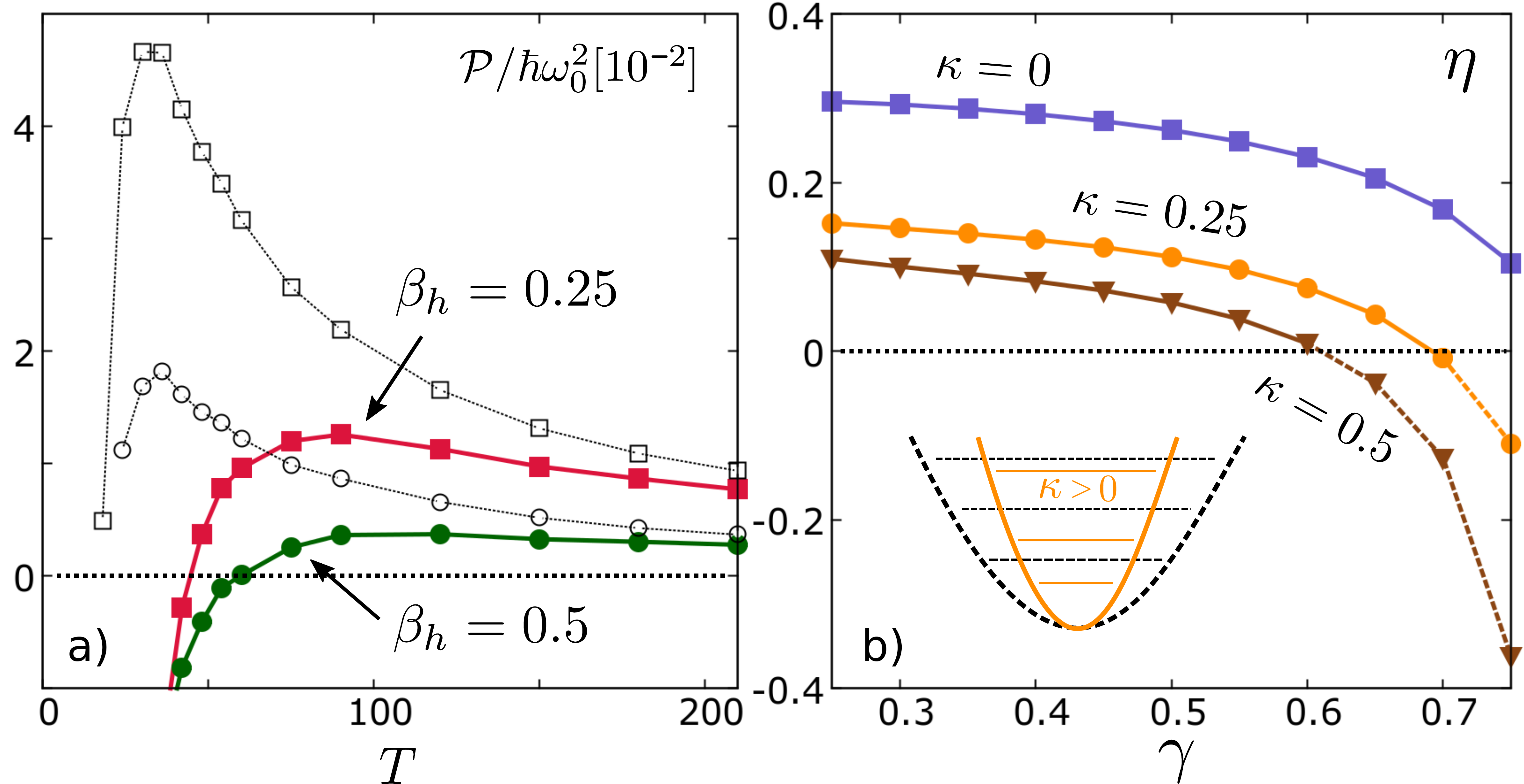}
 		\caption{\label{fig:power}  (a) Engine net output power $\mathcal P = -(W_d+W_I)/T$ as a function of the period $T$ for $\omega_0\beta_c = 3$, $\gamma/\omega_0 = 0.5$ and time scales $\tau_I = T/6$, $\tau_{d}=T/12$ compared to the respective output power ignoring $W_I$ (black dotted) and (b) efficiency vs. $\gamma$ for various anharmonicity parameters $\kappa$; here $\omega_0\tau_I = \omega_0\tau_{d} = 5$, $\omega_0T = 40$ and other parameters are as in Fig. \ref{fig:populations}.}
 		\end{figure}

As expected, values obtained for $\eta$ are always below the Curzon-Ahlborn and the Carnot efficiencies \cite{SM}, but yet, even beyond weak coupling, they do exceed $\eta\sim 0.2$. The coupling work appears as an essential ingredient also to predict for the power output the optimal cycle time and correct peak height, Fig.~\ref{fig:power}(a). If it is ignored, misleading data are obtained. Beyond the harmonic case, i.e. for stiffer anharmonic potentials, dynamical features discussed in Fig.~\ref{fig:populations}(d)  reduce the efficiency, Fig.~\ref{fig:power}b. They have a similar impact as  enhanced thermal couplings, both having the tendency to suppress (increase) fluctuations in position (momentum).

\begin{figure}[t]
	\centering
	    \includegraphics[width =\columnwidth]{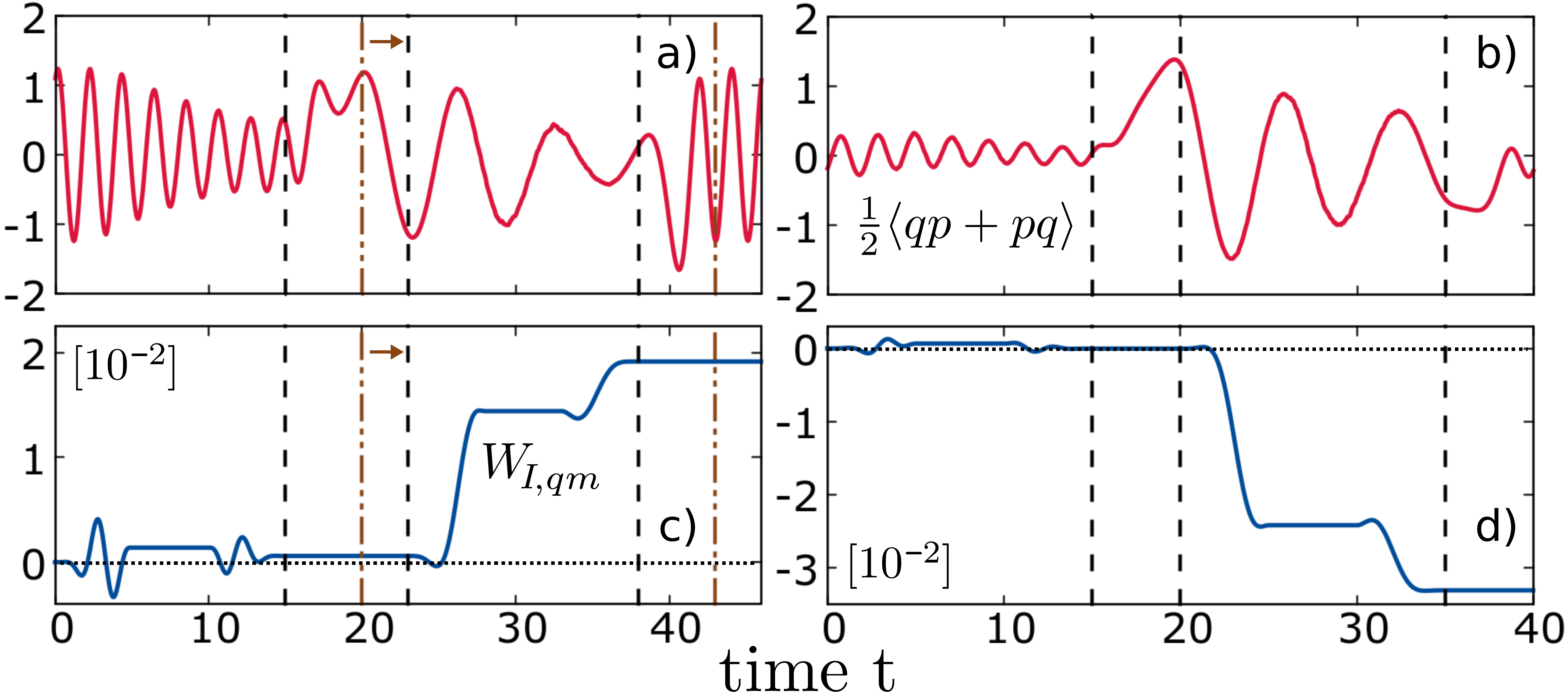}
		\caption{\label{fig:qpworksupp} The quantum part $W_{I, qm}$ of the coupling work explicitly depends on the phase of $qp$-correlations during the de-/coupling procedure. In (a, c) a segment is added to allow after the expansion stroke for a unitary time evolution at fixed frequency so that $W_{I, qm}>0$ while $W_{I, qm}<0$ for the standard  (b, d). Parameters are $\omega_0\hbar\beta_h = 0.25$, $\omega_0\hbar\beta_h = 3$, $\gamma/\omega_0 = 0.1$; other process time scales are as in Fig. \ref{fig:fullcycle}.}
\end{figure}

In conclusion, by simulating non-perturbatively and within a systematic formulation the dynamics of quantum thermal machines with single degrees of freedom as work medium, we have obtained a complete characterization of their properties. The medium-reservoir boundary appears as an internal feature of the model so that full control over the medium as well as its thermal contact to reservoirs is possible. The example of the four stroke Otto engine demonstrates the decisive role of the coupling work that must be considered as an integral part of the total energy balance. Its overall impact is detrimental to the efficiency of quantum heat engines, however, can be reduced by quantum correlations if they are properly controlled. A simple example 
is shown in Fig.~\ref{fig:qpworksupp}, where a slight change in the expansion stroke of the Otto engine's protocol modifies the phase of $qp$-correlations such that the opposite happens: coupling work is further enhanced and the efficiency further suppressed. This sensitivity of quantum heat engines to changes in the driving protocol can be exploited by optimal control techniques in future devices. The presented approach provides the required tools to follow theoretically these activities.

We thank R. Kosloff, E. Lutz, M. M{\"o}tt{\"o}nen, and J. Pekola for fruitful discussions. Financial support from the Land BW through the LGFG program (M.W.), the IQST and the German Science Foundation (DFG) through AN336/12-1 (J.A.) are gratefully acknowledged.

\bibliography{engineMJJ}{}

\end{document}


\title{- Supplemental material -\\ Out-of-equilibrium operation of a quantum heat engine:\\ The cost of thermal coupling control}

\author{Michael Wiedmann}
\author{J\"urgen T. Stockburger}
\author{Joachim Ankerhold}
\affiliation{Institute for Complex Quantum Systems and IQST, University of Ulm, 89069 Ulm, Germany}

\date{\today}

\begin{abstract}
We indicate how to adapt the SLN simulation method to the case of reservoirs with time-dependent coupling. This results in a complete, exact dynamical simulation of the quantum Otto engine in terms of the full, interacting dynamics of the work medium and its two thermal and harmonic reservoirs. We also provide expressions for work and heat, relating expectation values of the full system-plus-reservoir dynamics to SLN-based expressions. Some more detailed numerical results are presented and discussed.
\end{abstract}

\pacs{Valid PACS appear here}
\maketitle

\onecolumngrid
\section{\label{OQSdynamics} Stochastic mapping of open system dynamics with two reservoirs and driving}

We consider a distinguished quantum system $H_{m} (t) = \frac{p^2}{2m} + V(q,t)$ which is interacting with two thermal, harmonic reservoirs $H_{c/h}$ under influence of time-dependent bilinear coupling $H_{I, c/h}(t)=-\lambda_{\rm c/h}(t) q\sum_k c_{k, c/h} (b_{k, c/h}^\dagger+b_{k,c/h})+\frac{1}{2} q^2 \lambda_{\rm c/h}^2 (t) \mu_{\rm c/h}$
\begin{equation}
    H(t)=H_m(t)+  H_{\rm c} + H_{I, {\rm c}}(t) +  H_{\rm h} + H_{I, {\rm h}}(t).
    \label{eq:Htot}
\end{equation}
Starting from factorizing initial conditions for the global density matrix $\rho_{tot}$ and reservoirs with ohmic characteristics, i.e. spectral densities of the form $J(\omega) = m\gamma\omega/(1+\omega^2/\omega_{\text{cut}}^2)^2$ up to a high frequency cutoff $\omega_{\text{cut}}$ (significantly larger than any other frequency of the problem, including $1/\hbar\beta_{c/h}$) and potential renormalization $\mu = \frac{2}{\pi}\int_0^{\infty} d\omega \frac{J(\omega)}{\omega}$, the Feynman-Vernon path integral formulation for the reduced density operator of the medium can be converted into the Stochastic Liouville-von Neumann equation with dissipation (SLN) of the form~\cite{stock99} 
\begin{eqnarray}
\frac{d}{dt}\rho_{\xi}(t) &=& \frac{1}{i \hbar}[H_m(t),\rho_{\xi}] + \sum_{\alpha=c,h} \left\{ \frac{i}{\hbar} \lambda_{\alpha}(t)\xi_{\alpha}(t)[q,\rho_{\xi}] - i \frac{m\gamma_{\alpha}}{2 \hbar^2} \lambda_{\alpha}(t)\dot{\lambda}_{\alpha}(t) [q^2,\rho_{\xi}] \right.\nonumber \\
&& \left. - i \frac{m\gamma_{\alpha}}{2\hbar^2} \lambda_{\alpha}^2(t) [q, \{p,\rho_{\xi}\}] - \lambda_{\alpha}^2(t) \frac{m\gamma_{\alpha}}{\hbar^2 \beta_{\alpha}} [q,[q,\rho_{\xi}]]\right\}.
\label{sled+}
\end{eqnarray}
This procedure involves mapping the reduced system evolution to a stochastic propagation in probability space that is governed by two Gaussian noise sources $\xi_h(t)$, $\xi_c(t)$ that are constructed to match the real part of the respective reservoir correlation function $\langle \xi_{\alpha}(t)\xi_{\alpha}(t')\rangle = \Re L_{\alpha}(t-t') - \frac{2m\gamma_\alpha}{\hbar\beta_\alpha}\delta(t-t'),\,  \alpha=c, h$. The subtracted white-noise term is treated separately, leading to the last term in (\ref{sled+}).
In the case of ohmic reservoirs, the limit of large $\omega_{\text{cut}}$ also makes the dynamics response of the reservoir near instantaneous, allowing its representation by a Dirac delta term of the form $\Im L_{\alpha}(t-t') = \frac{m\gamma_{\alpha}}{2}\frac{d}{d t}\delta(t-t'), \, \alpha=c, h$. An integration by parts transforms $\Im L_{\alpha}(\tau)$ into a delta function, with a boundary term removing the $q^2$ term in the coupling, and introducing terms dependent on momentum $p$ and the time derivative $\dot\lambda$.  A single stochastic sample obtained by propagating the density with an individual trajectory possesses no direct physical meaning, except in the classical limit $\hbar\to 0$. The physical density matrix is obtained by averaging over a sufficiently large number of samples $\rho_m(t)=\mathbb{E}[\rho_\xi(t)]$.

Taking the spectral density of the reservoirs to be a smooth function of frequency implies the limit of an infinite number of environmental modes, with a correspondingly infinitesimal coupling of each individual mode. This procedure also gives the reservoirs infinite heat capacity, allowing them to be treated as thermodynamic reservoirs even if only their initial state is specified as thermal.

\section{\label{HEcycle} Heat engine cycle}
We will now consider a four stroke quantum heat engine with an harmonic or an anharmonic oscillator as working medium. Our description is based on a non-perturbative propagation of the reduced density matrix within the dissipative SLN probabilistic framework. The model comprises three control parameters. The hot and cold baths are controlled with the time-dependent coupling parameters $\lambda_{c/h}(t)$, the oscillator potential is frequency modulated through $\omega(t)$
\begin{equation}
    V(q,t)=\frac{1}{2} m \omega^2(t) q^2 + \frac{1}{4} m \kappa q^4.
\end{equation}
\subsection{Energy balance and first law}
We first consider the energy balance of our engine in the context of the full system-reservoir model.
Any changes in the energy of the global system as defined in Eq. (\ref{eq:Htot}) are due to work terms,
\begin{equation}
\frac{d}{dt}\langle H \rangle =\left\langle \frac{\partial H_m(t)}{\partial t}\right\rangle
+ \left\langle \frac{\partial H_{I,c}(t)}{\partial t}\right\rangle
+ \left\langle \frac{\partial H_{I,h}(t)}{\partial t}\right\rangle,
\end{equation}
with separate terms indicating the different modes of performing work associated with the parameter $\omega(t)$ and $\lambda_{c/h}(t)$. Heat is identified as energy transferred into the reservoirs over a cycle,
\begin{equation}
    Q_{c/h} := - \int\limits_0^T dt \text{Tr}\{H_{c/h}\dot \rho_{tot}\}
    = -\frac{i}{\hbar} \int\limits_0^T dt \text{Tr} \{[H(t),H_{c/h}]\rho_{tot}\}
    = -\frac{i}{\hbar} \int\limits_0^T dt \text{Tr} \{[H_{I,c/h}(t),H_{c/h}]\rho_{tot}\}.
    \label{eq:Qchdef}
\end{equation}
At the end of a period of cyclic operation, the microstate of the medium reverts its initial state, $\rho_m(t+T) = \rho_m(t)$. Moreover, the collective response function of the reservoirs decays in time sufficiently fast that the collective reservoir coordinate entering $H_{I,c/h}$ shows periodic behavior at long times. We thus have
\begin{equation}
    \langle H_m(t+T)\rangle = \langle H_m(t)\rangle,\; \langle H_{I,c/h}(t+T)\rangle = \langle H_{I,c/h}(t)\rangle.
    \label{eq:periodic}
\end{equation}
Defining per-cycle work terms through
\begin{equation}
    W_d = \int_0^T dt \left\langle \frac{\partial H_m(t)}{\partial t}\right\rangle,\;
    W_I = \int\limits_0^T dt \left( \left\langle\frac{\partial H_{I,h}}{\partial t}\right\rangle
    + \left\langle\frac{\partial H_{I,c}}{\partial t}\right\rangle \right)\, ,
\end{equation}
it is easily verified that these quantities obey the first law of thermodynamics
\begin{equation}
    W_d + W_I + Q_c + Q_h = 0\, .
\end{equation}
It is instructive to note that Eq. (\ref{eq:periodic}) can be used to give alternative expression for the heat terms per period,
\begin{equation}
    Q_{c/h} = \frac{i}{\hbar} \int\limits_0^T dt \text{Tr} \{[H_{I,c/h}(t), H_m]\rho_{tot}\}
    - \int\limits_0^T dt \left\langle\frac{\partial H_{I,c/h}}{\partial t}\right\rangle
    \label{eq:Qch2}
\end{equation}
The first term of (\ref{eq:Qch2}) is an energy flow from the system due to the coupling; the second is the work performed through changes of $\lambda_{c/h}$. Viewing this equation as a continuity equation, we conclude that the work described by the second term is completely dissipated.

\subsection{Work and heat in the probabilistic SLN context}

The work terms $W_d$ and $W_I$ as well as the expression (\ref{eq:Qch2}) for the heat transfer $Q_{c/h}$ have equivalent expressions in the SLN dynamics (\ref{sled+}), even for those terms involving $H_{I,c/h}$. In order to obtain this stochastic equivalent, the equal-time correlations $\langle q X_\alpha\rangle$ and $\langle p X_\alpha\rangle$, ($\alpha=c, h$), involving system coordinate/momentum and the reservoir operator $X_{\alpha}=\sum_k c_{k, \alpha} (b_{k, \alpha}^\dagger+b_{k,\alpha})$ are needed. In simpler SLN approaches~\cite{stock02}, the noise variable $\xi$ can serve as a direct substitute for $X$~\cite{schmi15}.
Here we need to treat the terms which have been contracted to delta terms in the more complicated SLN equation (\ref{sled+}) separately. A careful consideration of short-time dynamics on timescales of order $1/\omega_{cut}$ before taking the limit of large $\omega_{cut}$ leads to the results
 \begin{align}
     \text{Tr}\{ q X_\alpha \rho_{tot}\} &= \mathbb{E} \left[\xi_\alpha(t)\langle q\rangle + \lambda_\alpha(t)\mu_\alpha \langle q^2\rangle
      - \lambda_\alpha(t) \gamma_\alpha\langle qp + pq\rangle/2 - \dot\lambda_\alpha(t) m \gamma_\alpha \langle q^2\rangle
     \right]\\
     \text{Tr}\{ p X_\alpha \rho_{tot}\} &= \mathbb{E} \left[\xi_\alpha(t)\langle p\rangle + m\gamma_\alpha \lambda_\alpha(t) k_{B}T_\alpha
     + \lambda_\alpha(t)\mu_\alpha \langle qp + pq\rangle/2 - \lambda_\alpha(t) \gamma_\alpha \langle p^2\rangle - \dot\lambda_\alpha(t)m\gamma_\alpha \langle qp+pq \rangle/2 
     \right]
 \end{align}
where $\alpha=c, h$. Table \ref{tab:formulae} summarizes expressions for work and heat using either the full Hamiltonian dynamics or the SLN framework. All quantities appearing in the SLN column can be extracted from simulation data. The expression for $Q_{c/h}$ is based on Eq. (\ref{eq:Qch2}), thus avoiding expressions involving momenta or velocities of the reservoir.
\begin{table}
\begin{tabular}[b]{|p{1cm}||l|l|}
\hline
     & Hamiltonian dynamics & SLN dynamics  \\ \hline \hline
    work & $\: W_d = \int_0^T dt \langle \frac{\partial H_m(t)}{\partial t}\rangle$
    & $\: W_d = \mathbb{E}\left[\int_0^T dt \:\omega(t)\dot{\omega}(t) \langle q^2 \rangle\right] $ \\[1ex]
    \cline{2-3}
    & 
    
    $\begin{array}[b]{r@{}l@{}}
    W_I & {}= \int_0^T dt \left[ \langle\frac{\partial H_{I,h}}{\partial t}\rangle + \langle\frac{\partial H_{I,c}}{\partial t}\rangle \right]\\ {}& 
    \end{array}$
    & 
    $\begin{array}[b]{r@{}l@{}}
W_I = \mathbb{E} \Big[& {}\int_0^T dt \left\{ - \left( \dot{\lambda}_h(t)\xi_h(t)+\dot{\lambda}_c(t)\xi_c(t) \right) \langle q \rangle\right. \nonumber \\
&\left.+\left(  \gamma_h \lambda_h(t)\dot{\lambda}_h(t)+ \gamma_c \lambda_c(t)\dot{\lambda}_c(t) \right)  \langle qp +pq\rangle/2\right.\nonumber \\
&\left.+\left(  \gamma_h \dot{\lambda}_h^2 + \gamma_c \dot{\lambda}_c^2 \right) m\langle q^2 \rangle\right\}\Big]\nonumber
    \end{array}$
    \\[1ex] \hline \hline 
    $\begin{array}[b]{r@{}l@{}}
    \text{heat}\\
    \text{\phantom{}}\\
    \text{\phantom{}}\\
    \text{\phantom{}}
    \end{array}$
    &  $\begin{array}[b]{r@{}l@{}}
       Q_{c/h} &{}= - \int_0^T dt \text{Tr}\{H_{c/h}\dot \rho_{tot}\}\\
    &{}= -\frac{i}{\hbar} \int_0^T dt \langle[H_{I,c/h}(t),H_{c/h}]\rangle \\ &{}
    \end{array}$
    & 
$\begin{array}[b]{r@{}l@{}}
    Q_{c/h} =
         \mathbb{E}\Big[ &{} \int_0^T dt \Big\{\lambda_{c/h}(t)\xi_{c/h}(t)\langle p \rangle / m\\ 
         &-\gamma_{c/h}\lambda_{c/h}^2(t) \langle p^2\rangle /m
         + \gamma_{c/h}\lambda_{c/h}^2(t) k_B T_{c/h} \\
         &-\gamma_{c/h}\lambda_{c/h}\dot\lambda_{c/h}\langle qp+pq \rangle\\
         &+ \dot{\lambda}_{c/h}(t)\xi_{c/h}(t)\langle q \rangle 
         - \gamma_{c/h} \dot{\lambda}_{c/h}^2(t)m \langle q^2\rangle 
          \Big\}\Big]
         \end{array}$ \\[1ex]
    \hline
\end{tabular}
\caption{\label{tab:formulae}Work and heat expressed in terms of the unitary evolution of system and reservoir and in terms of SLN propagation of a periodic steady state.}
\end{table}

\subsection{Estimates for driving and coupling work}

We start from the formulation of the driving and the coupling work in the SLN context as specified in Table~\ref{tab:formulae}. The driving work provides contributions along the isentropic strokes of length $\tau_d$ with $\dot{\omega}(t)<0$ during B$\to$ C and $\dot{\omega}(t)>0$ during D$\to$ A. The coupling work provides contributions within time intervals of length $\tau_I$ at the beginning ($\dot{\lambda}_\alpha>0$) and the end ($\dot{\lambda}_\alpha<0$) of the isochoric strokes. 

To estimate the above integrals we now assume the following: (i) $\dot{\omega}(t)=const.$ during $\tau_d$ and $\dot{\lambda}(t)=const.$ during $\tau_I$, (ii) the contribution of the variances dominates in $W_I$ and (iii) within $\tau_d$ the variances change linearly for $\Delta\omega/\omega_0\ll 1$. 

We then obtain
\begin{eqnarray}
W_d &\approx& m|\dot{\omega}| \frac{\omega_0\tau_d}{3} \left(\langle q^2\rangle_A \Omega_++\langle q^2\rangle_D \Omega_- -\langle q^2\rangle_C \Omega_- -\langle q^2\rangle_B \Omega_+\right)\nonumber\\
&\approx& m\dot{\omega} \frac{\omega_0\tau_d}{2} \left(\langle q^2\rangle_A+\langle q^2\rangle_D-\langle q^2\rangle_C -\langle q^2\rangle_B\right)\, ,
\end{eqnarray}
where $\Omega_\pm=3/2\pm (\Delta\omega/2\omega_0)\approx 3/2$ and $\langle q^2\rangle_X=\mathbb{E}[\langle q^2(t_X)\rangle]$ with $t_X$ corresponding to point $X$ in the cycle.  The coupling work can be approximated by
\begin{equation}
W_I\approx m\gamma\tau_I \dot{\lambda}^2 \left(\langle q^2\rangle_A+\langle q^2\rangle_D+\langle q^2\rangle_C+\langle q^2\rangle_B\right)\, ,
\end{equation}
where we considered $\langle q^2\rangle \approx const $ during switching the coupling off or on for sufficiently short $\tau_I$. 

Due to the general relations 
\begin{equation}
\langle q^2\rangle_C > \langle q^2\rangle_D\ \mbox{(contact to cold bath)}\ , \  \langle q^2\rangle_B > \langle q^2\rangle_A\ \mbox{(contact to hot bath)}
\label{relation}
\end{equation}
 for $\gamma>0$, one first concludes that $W_d<0$ while apparently $W_I>0$. Further, one finds for the total net work
\begin{equation}
W_d+W_I\approx a_- \left(\langle q^2\rangle_C+\langle q^2\rangle_B\right) + a_+\left(\langle q^2\rangle_A+\langle q^2\rangle_D\right)
\end{equation}
with $a_\pm = m\gamma\tau_I \dot{\lambda}^2\pm m\tau_d \omega_0|\dot{\omega}|/2$. Now, to qualify for a heat engine, the condition $W_d+W_I<0$ needs to be fulfilled which implies
\begin{equation}
\frac{\omega_0\tau_I\Delta\omega-2\gamma}{\omega_0\tau_I\Delta\omega+2\gamma}>R\equiv \frac{\langle q^2\rangle_A+\langle q^2\rangle_D}{\langle q^2\rangle_C+\langle q^2\rangle_B}\, 
\end{equation}
with $\tau_d|\dot{\omega}|=\Delta\omega$ and $\tau_I\dot{\lambda}^2=1/\tau_I$. Due to (\ref{relation}) one always has $0\leq R\leq 1$. The above relation can be easily solved for $\tau_I$ to read
\begin{equation}
\omega_0\tau_I> \frac{2\gamma}{\Delta\omega}\frac{1+R}{1-R}
\end{equation}
as a condition on the minimal time $\tau_I$ consistent with heat engine operation.\\

Several situations can now be considered analytically:

\begin{enumerate}
  \renewcommand{\labelenumi}{(\roman{enumi})}
 
  \item $\gamma\to 0$: In this limit a perturbative treatment in $\gamma$ applies. The equations of motions for the cumulants then tell us that deviations from the unitary time evolution are on order $O(\gamma)$ which implies $R=1-O(\gamma)$. Hence, $(1-R)/\gamma\approx const.$ and the minimal  $\tau_I$ approaches a constant value.
 
 \item In the classical regime (high temperature) we assume quasi-equilibrium throughout the cycle with $\langle q^2\rangle_A=1/(m\beta_c\omega_h^2)$ etc. which yields $R_{\rm cl}=T_c/T_h$.
 
 \item In the quantum regime (lower temperatures) in quasi-equilibrium one has $R>R_{\rm cl}$.
 
 \item For low temperatures and sufficiently large $\gamma\tau_R$ to allow for states close to thermal equilibrium, the variances in position depend only weakly on $\gamma$ for $\gamma\hbar\beta<2\pi$. Accordingly, the $\gamma$-dependence of the threshold of $\tau_I$ is predominantly given by $\gamma/\Delta\omega$.
 
 \item For $\hbar\beta_c\omega_c\gg 1$ while $\hbar\beta_h\omega_h\ll 1$ and  $\gamma\tau_R>1$ to allow for approximate equilibration  with $\langle q^2\rangle_A\approx \hbar/m\omega_h$, $\langle q^2\rangle_C\approx 1/m\omega_h^2\beta_h$ etc., one finds $R\approx \hbar\beta_h\omega_0$. For the parameters in Fig.~4 of the main text we then have $\omega_0\tau_I > 12 \gamma/\omega_0$ which, for the given value of $\tau_R$, describes the minimal $\tau_I$ sufficiently accurate for all $\gamma/\omega_0>0.1$.
\end{enumerate}

\subsection{Quantum effects in the coupling work}

As seen in Table \ref{tab:formulae}, the coupling work consists of the contributions $W_I=W_{I, cl}+W_{I, qm}$. The first classical-like part is determined by the variance $\langle q^2\rangle$, the quantum part is determined by the $\xi \langle q\rangle$ average and the contribution of the coherences $\langle qp+pq\rangle$. In the high temperature limit, only the $W_{I, cl}$ part survives so that the remaining two contributions represent a genuine quantum effect. The quantum contribution $W_{I, qm}$ is dominated by the part depending on the $qp$-correlations. It is this part which we want to consider in the following in more detail:
\begin{equation}
W_I^{(qp)}=\frac{\gamma}{2} \int_0^T dt \left(\lambda_h(t) \dot{\lambda}_h(t) \langle qp+pq\rangle+\lambda_c(t) \dot{\lambda}_c(t) \langle qp+pq\rangle\right)\, .
\end{equation}
Now, following the above arguments we assume $\lambda_\alpha(t)=t/\tau_I$ when the coupling is switched on, and $\lambda_\alpha=(1-t/\tau_I)$ when the coupling is switched off. Further we assume that
\begin{eqnarray}
\langle (qp+pq)(t)\rangle &\approx &\langle qp+pq\rangle_A \, \cos(2\omega_h t) {\rm e}^{-\gamma t}\ \ \mbox{(switching on hot isochore)}\nonumber\\
\langle (qp+pq)(t)\rangle &\approx &\langle qp+pq\rangle_A \, \cos[2\omega_h (\tau_I+\tau_R+t)] {\rm e}^{-\gamma (\tau_I+\tau_R+t)}\ \ \mbox{(switching off hot isochore)}
\end{eqnarray}
where $0\leq t\leq \tau_I$ and $\omega_h=\omega_0+\Delta\omega/2$. For the cold isochore we replace $A\to C$ and 
$\omega_h\to\omega_c=\omega_0-\Delta\omega/2$.

This then allows us to estimate  $W_I^{(qp)}$ as
\begin{equation}
W_I^{(qp)}\approx \frac{\gamma}{2\tau_I}\left( \langle qp+pq\rangle_A \, \Omega_h+\langle qp+pq\rangle_C\, \Omega_c\right)\, ,
\end{equation}
where
\begin{equation}
\Omega_\alpha=\int_0^{\omega_0 \tau_I} dx \left\{\frac{x}{\omega_0\tau_I} \cos\left(2 \frac{\omega_\alpha}{\omega_0} x\right){\rm e}^{-(\gamma/\omega_0)x}-\left(1-\frac{x}{\omega_0\tau_I}\right) \cos\left[2 \frac{\omega_\alpha}{\omega_0} (\omega_0 \tau_I+\omega_0\tau_R+x)\right]{\rm e}^{-(\gamma/\omega_0)(\omega_0 \tau_I+\omega_0\tau_R+x)}\right\}\, .
\end{equation}
These latter dimensionless expressions are functions oscillating with varying $\tau_I$ for fixed other parameters with the typical behavior that $|\Omega_c|>|\Omega_h|$ when $\omega_0\tau_I$ is not too small. Apparently, they exhibit strong (weak) oscillations for hot (cold) ($\omega_h>\omega_c$)  isochores. 

For $\gamma (\tau_I +\tau_R)>1$, one finds for the above integral
\begin{equation}
W_I^{(qp)}\approx  -\frac{\omega_0}{\tau_I \omega_\alpha^2}\left[1+{\rm e}^{-\gamma\tau_I}(\gamma\tau_I-1) \cos(2\omega_\alpha\tau_I)+\omega_0 \tau_I\sin(2\omega_\alpha\tau_I)\right]\, .
\end{equation}

Consequently, the sign of $W_I^{(qp)}$ depends on $\omega_\alpha\tau_I$ but also on the phase of the $qp$-correlations. We find the dominant contribution to $W_I^{(qp)}$ is provided by $\Omega_c \langle qp+pq\rangle_C$, i.e. after expansion and before coupling to the cold reservoir. For the parameters chosen in Fig.~3 in the main text, one always has $\langle qp+pq\rangle_C>0$ so that $W_{I, qm}$ counteracts $W_{I, cl}$ to lead to a reduced $W_I$.

However, as seen in the Fig.~6 of the main text, by extending the unitary time evolution after expansion/compression such that the system evolves for about half a period $\pi/\omega_\alpha$ at constant frequency, the phase of the $qp$-correlations at $C$ turns from positive to negative. Hence, $W_{I, qm}>0$ and adds to $W_{I, cl}$ to enhance $W_I$.

\subsection{\label{numerics}Details about the numerical implementation}
The numerical solution of the dissipative SLN leads to one single trajectory of the reduced system density. This is realized by moving to position representation using symmetric and antisymmetric coordinates, i.e. $\rho_{\xi}(r,y) = \langle r-\frac{y}{2} |\rho_\xi |r+\frac{y}{2}\rangle$. This representation allows an efficient split-operator technique. In addition to the commonly used FFT method (alternating between diagonal potential and kinetic terms) we employ a third step related to the operator $[q,\{p,\cdot\}]$, which can be understood to be the generator of a re-scaling operation. For typical parameters, e.g., parameters $\omega_0\hbar\beta_h = 0.25$, $\omega_0\hbar\beta_c = 3$, $\omega_{\text{cut}}/\omega_0 = 30$, $\gamma/\omega_0 = 0.25$ and characteristic cycle time scales $\omega_0\tau_I = 10$, $\omega_0\tau_d = \omega_0\tau_{R} = 5$, $\omega_0T = 60$ up to the first three cycles of a  periodic steady-state takes approximately $72$ CPU core hours on a Intel Xeon CPU (Sandy Bridge architecture).  A typical number of samples $n_{samp} \approx 500$; the resulting statistical errors are less than the line width or symbol size used in our figures.

\subsection{\label{addresults}Additional results}

Here we show additional results that are not included in the main text but provide further indications on the features and versatility of our simulation platform.\\

\begin{figure}[ht]
	\centering
		\includegraphics[width =0.75\columnwidth]{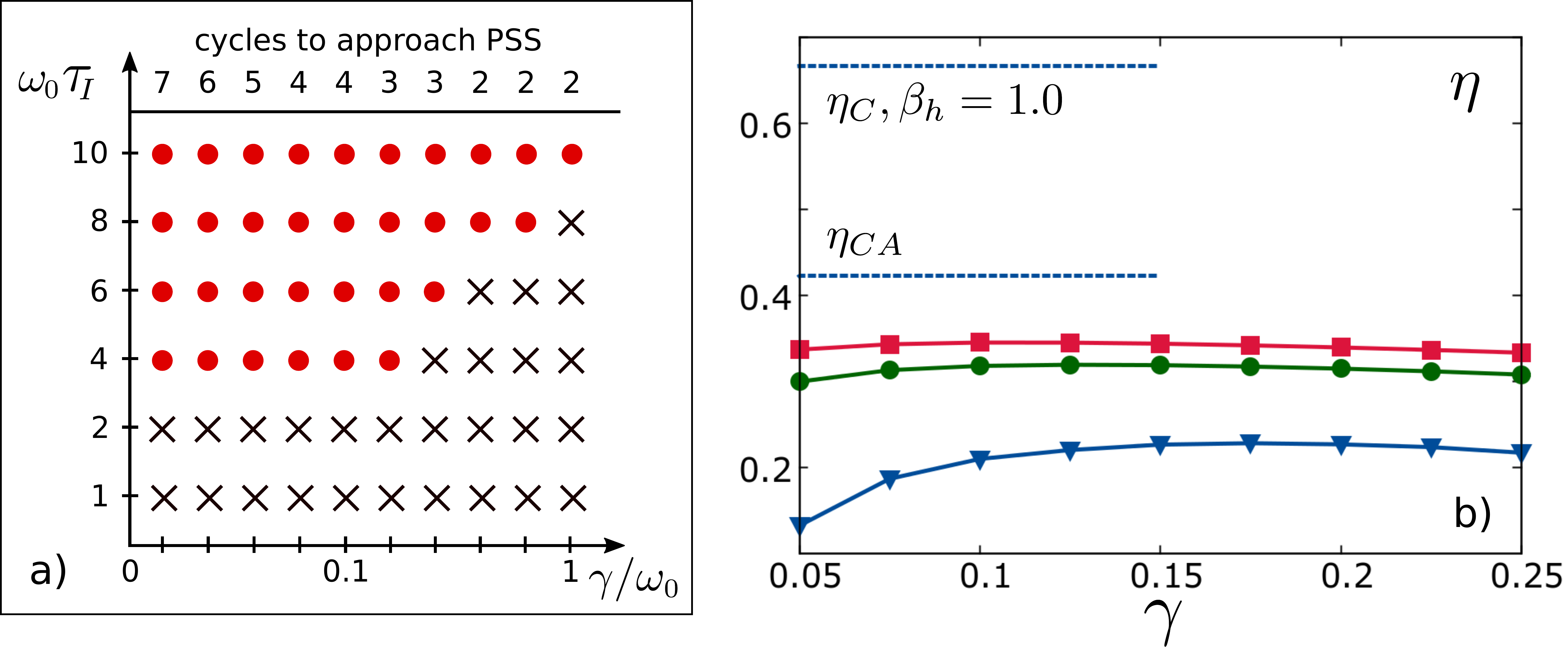}%
		\caption{\label{fig:harmeffsupp} (a) The PSS's operating regime as a QH (red circles) with efficiency $\eta>0$ vs.\ the maximal medium-reservoir coupling and the coupling time. Also shown is the transient time until PSS cycle is reached (top axis). Model parameters are $\omega_0\hbar\beta_h = 0.5 $, $\omega_0\hbar\beta_c = 3 $, $\omega_c/\omega_0 = 30$, $\omega_0\tau_{d}=\omega_0\tau_R=5$ and $T = 4 \tau_I +2\tau_d+ 2\tau_R$. (b) Harmonic heat engine efficiency vs. damping strength for hot bath temperatures $\omega_0\hbar\beta_h = 0.25$ (red), $\omega_0\hbar\beta_h = 0.5$ (green) and $\omega_0\hbar\beta_h = 1.0$ (blue) compared to Carnot and Curzon-Ahlborn efficiency for the coldest temperature. The cold bath is always fixed at $\omega_0\hbar\beta_h = 3$. Other model parameters are $\omega_0\tau_{d}=5$, $\omega_0\tau_I=10$ and $\omega_0 T = 60$.}
\end{figure}

Fig.~\ref{fig:harmeffsupp}a gives an alternative rendition of the PSS's operating regime as a QH, depending on the maximal medium-reservoir coupling and the coupling time as parameters. Numbers on top indicate the number of driving periods before the PSS cycle is reached. Fig.~\ref{fig:harmeffsupp}b compares the efficiency of the finite-time harmonic engine to the Carnot and the Curzon-Ahlborn efficiencies. The efficiencies lie always below the analytical values of $\eta_{CA}$ and $\eta_C$. For decreasing damping strength, the efficiencies tend to decrease as a result of diminished heat transfer and driving work. Stronger damping meanwhile fosters energy losses due to irreversible coupling work which also leads to decreased efficiency.\\ 

\begin{figure}[ht]
	\centering
		\includegraphics[width =0.5\columnwidth]{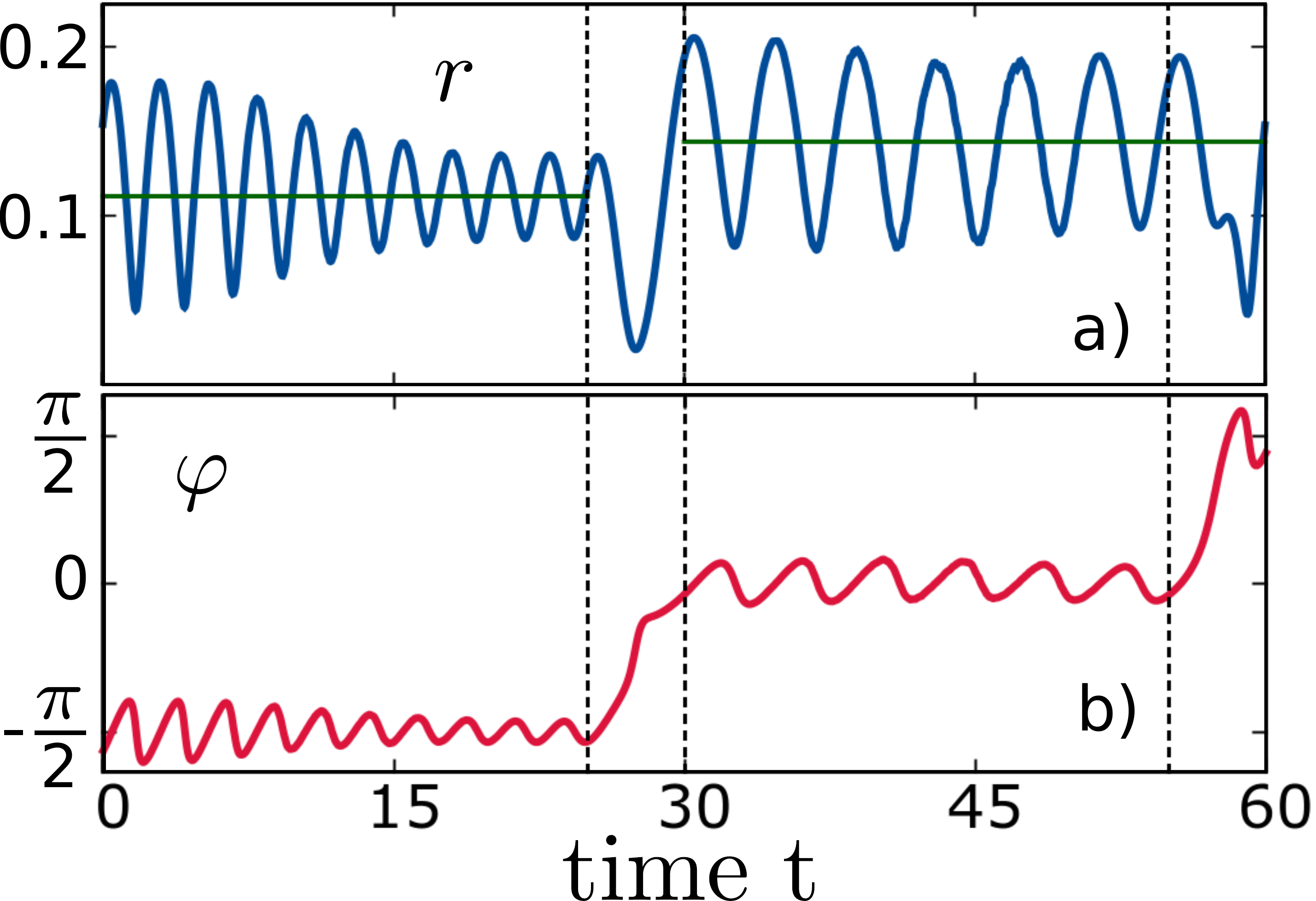}%
		\caption{\label{fig:squeeze} Quantum dynamics for an Otto engine with $\omega_0\hbar\beta_h = 0.25$, $\omega_0\hbar\beta_c = 3$. Time scales are $\omega_0\tau_I = 10$, $\omega_0\tau_{d} = 5$, $\omega_0 T = 60$ with reservoir coupling $\gamma/\omega_0 = 0.05$ and $\omega_{\rm cut}/\omega_0 = 30$. (a) Squeezing amplitude r compared to the values in thermal equilibrium for hot/cold baths (green), and (b) squeezing angle $\varphi$ for one cycle. The phase space ellipsoid makes a full turn ($2\varphi$ changes by $2\pi$) with its width oscillating around equilibrium values for hot/cold baths.}
\end{figure}

An alternative representation of the covariances is their parameterization through a squeezing amplitude r and a squeezing angle $\varphi$ \cite{weedb12}. In Fig.~\ref{fig:squeeze}a the squeezing amplitude r for the PSS operating regime of a QH is compared to the values in thermal equilibrium for hot/cold baths. While the amplitude is damped towards the equilibrium values in a damped oscillatory pattern, it maintains its non-equilibrium characteristics across a whole cycle. Fig.~\ref{fig:squeeze}b shows the movement of the phase space ellipsoid during the cyclic operation.\\

\begin{figure}[ht]
	\centering
		\includegraphics[width =0.75\columnwidth]{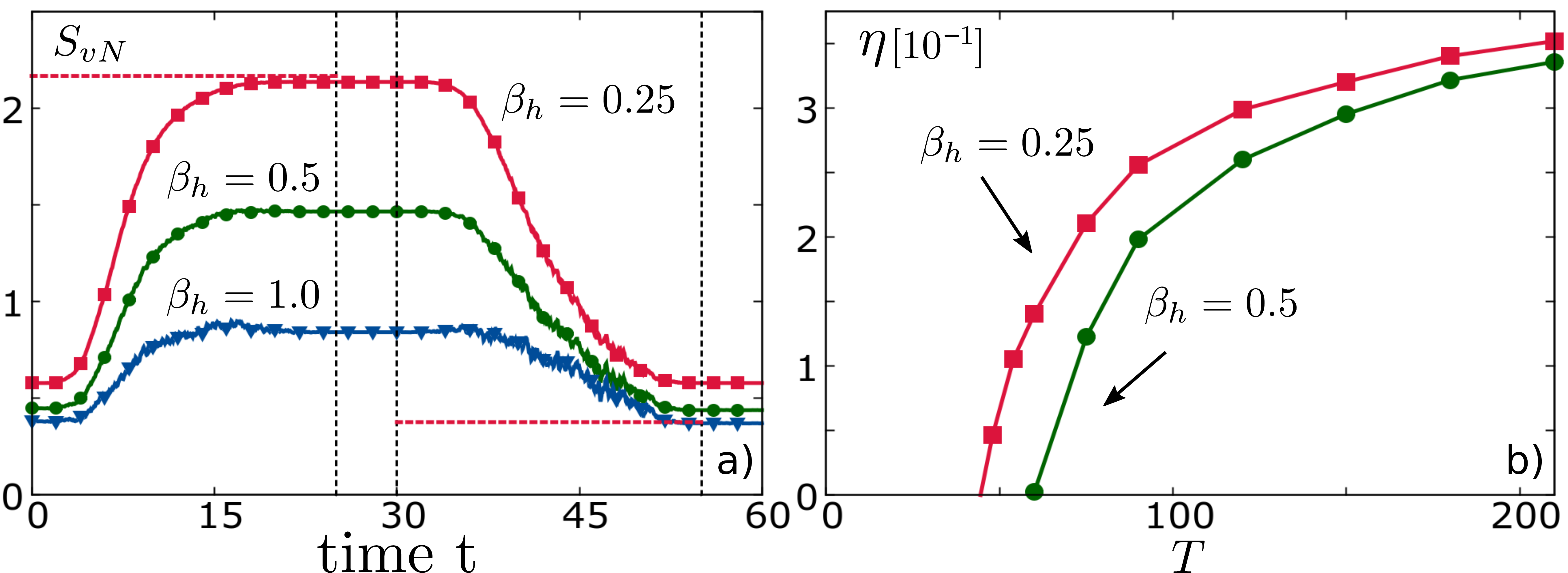}%
		\caption{\label{fig:entropyeffvar} (a) The von Neumann entropy $S_{vN}(\rho_m)$ for one cycle and various inverse hot bath temperatures $\beta_h$ at $\omega_0\hbar\beta_c = 3$ and $\gamma/\omega_0 = 0.25$ compared to the equilibrium entropy of hot ($\omega_0\hbar\beta_h=0.25$) and cold bath. (b) Engine efficiency as a function of the period $T$ for $\omega_0\beta_c = 3$, $\gamma/\omega_0 = 0.5$ and time scales $\tau_I = T/6$, $\tau_{d}=T/12$.}
\end{figure}

In the QH regime the PSS substantially deviates from a mere sequence of equilibrium states as also illustrated by the von Neumann entropy $S_{vN}$ in Fig.~\ref{fig:entropyeffvar}a. Even with a relatively long contact time compared to $1/\gamma$, the entropy values alone indicate incomplete thermalization with the colder reservoir, even when the non-thermal nature of squeezing is disregarded. Slowing the cycle in Fig.~\ref{fig:entropyeffvar}b tends to increase efficiency whereas maximum power peaks are significantly lowered and shifted when coupling work is no longer ignored (main text).\\ 

\begin{figure}[ht]
	\centering
		\includegraphics[width =0.7\columnwidth]{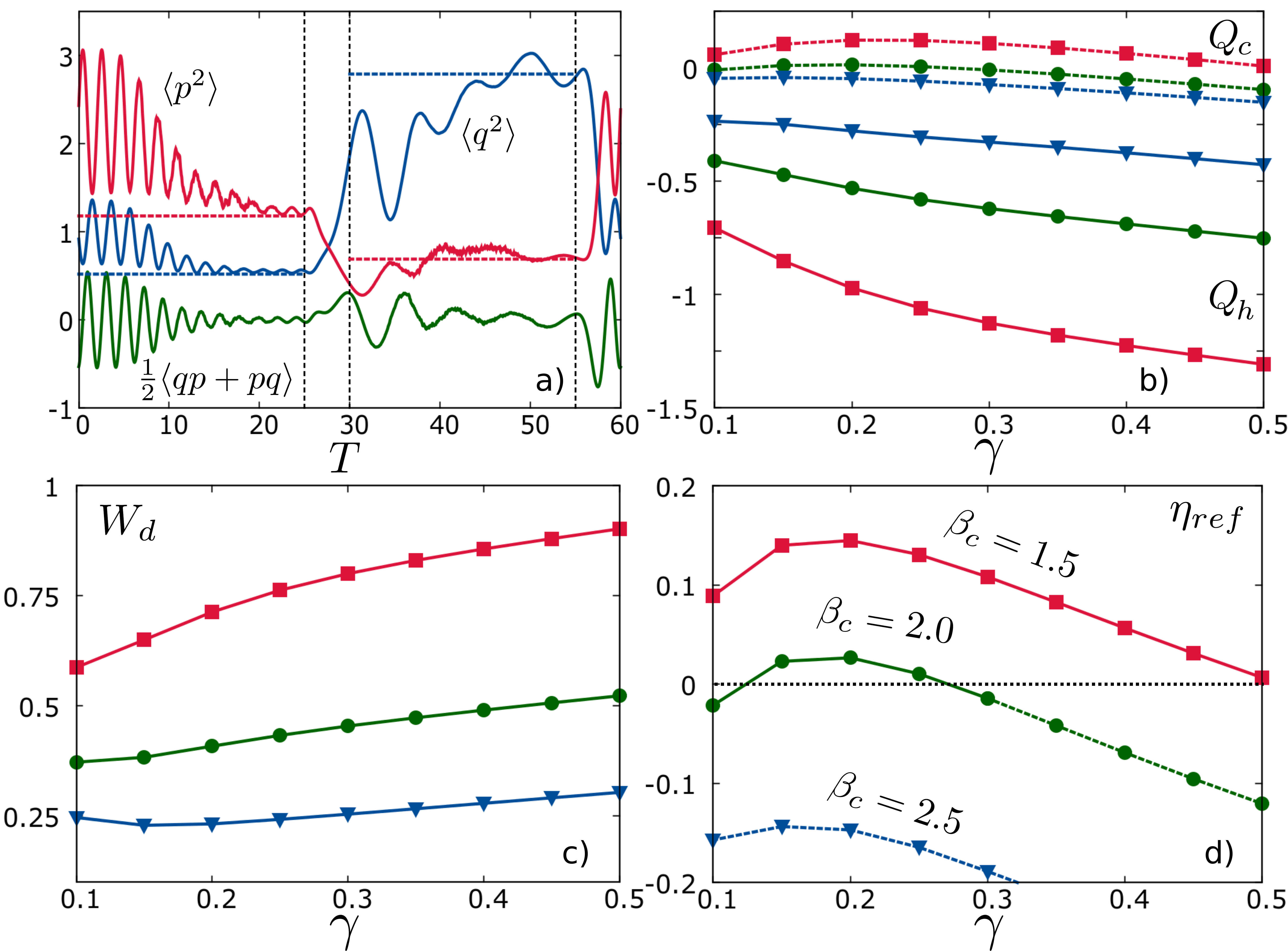}%
		\caption{\label{fig:refsupp}  Refrigerator setting: (a) Dynamics of the position $\langle q^2\rangle$ and momentum $\langle p^2\rangle$ dispersion and the symmetric mean $\frac{1}{2}\langle qp + pq\rangle$ for $\omega_0\hbar\beta_h = 10$, $\omega_0\hbar\beta_c = 1.5$ , $\omega_c/\omega_0 = 30$ and dissipation constant $\gamma/\omega_0 = 0.25$ compared to thermal equilibrium dispersions for one PSS cycle (b) heat that is absorbed from the cold reservoir $Q_c$ and released into the hot reservoir $Q_h$ (c) driving work $W_d$ and (d) formally determined efficiency $\eta_{ref} = Q_c/(W_d + W_I)$ versus $\gamma$; model parameters are $\omega_0\tau_{d}=5$, $\omega_0\tau_I=10$ and $\omega_0 T = 60$. }
\end{figure}

Fig.~\ref{fig:refsupp} shows details of the dynamics of the medium in a QR setting. These include (a) moments which fully characterize a non-thermal Gaussian state as well as work, heat, and efficiency for a refrigerator setting (b-d).\\

\begin{figure}[h]
	\centering
		\includegraphics[width =0.75\columnwidth]{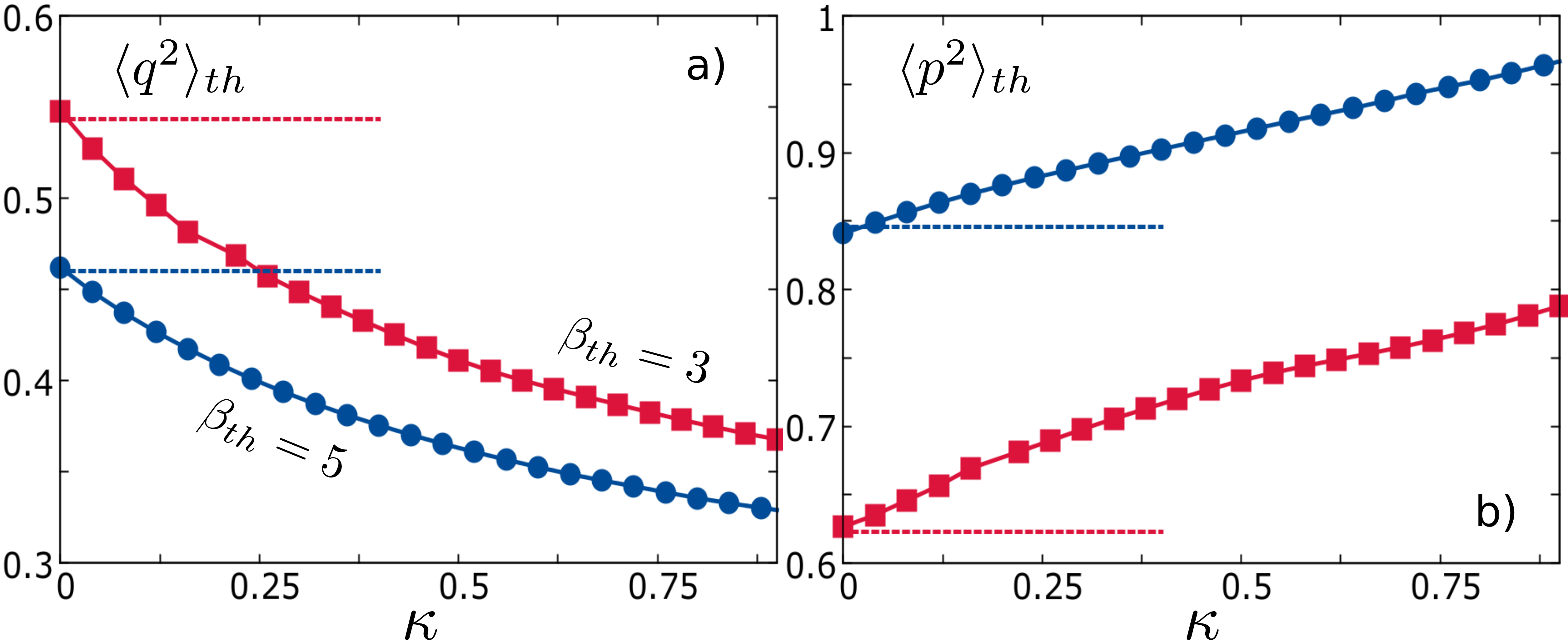}%
		\caption{\label{fig:anharmqp} Anharmonic oscillator in thermal equilibrium with one single thermal reservoir: equilibrium position $\langle q^2 \rangle_{th}$ and momentum $\langle p^2 \rangle_{th}$ dispersion for $\gamma/\omega_0 = 0.1$, $\omega_0\hbar\beta_{th} = 3$ (red) and $\gamma/\omega_0 = 0.5$, $\omega_0\hbar\beta_{th} = 5$ (blue) versus the anharmonicity $\kappa \in [0,0.9]$; with increasing $\kappa$, the oscillator potential gets stiffer and leads to decreasing second moments in position while those of the momentum are broadened; the horizontal lines correspond to analytical equilibrium values in the harmonic case.}
\end{figure}

Fig.~\ref{fig:anharmqp} shows variances in position and momentum in thermal equilibrium for anharmonic oscillators with varying anharmonicity parameter, illustrating the fact that the oscillator ``stiffness'' becomes effectively temperature dependent. The resulting deformations of equilibrium position and momentum variances have significant impact on heat, work and efficiency properties for increasing $\kappa$ (see discussion in the main text).\\

\begin{figure}[ht]
	\centering
		\includegraphics[width =0.5\columnwidth]{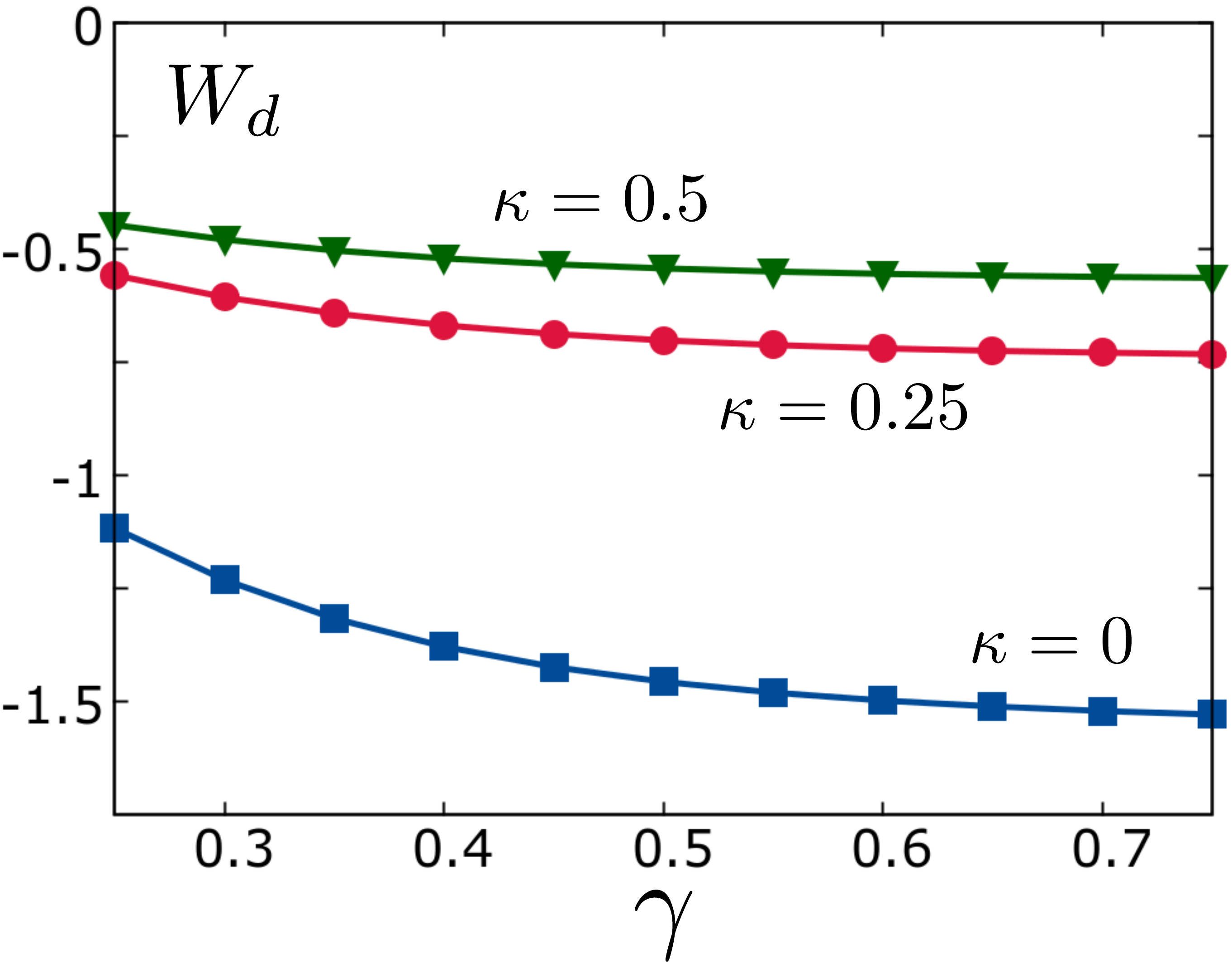}%
		\caption{\label{fig:workdrivesupp} Anharmonic setting: Driving work $W_d$ vs.\ $\gamma$ for various anharmonicity parameters $\kappa$, $\omega_0\tau_I = \omega_0\tau_{d} = 5$, $\omega_0T = 40$, $\omega_0\hbar\beta_h = 0.25$ and $\omega_0\hbar\beta_c = 3$. This reduction for stiffer potentials is determined by a reduced second moment $\langle q^2\rangle$ (see also Table \ref{tab:formulae}) as a result of the parametric drive. In terms of efficiency, anharmonicities seem then to play a similar role as enhanced thermal couplings, both having the tendency to localize the oscillator degree of freedom in position.}
\end{figure}

Fig.~\ref{fig:workdrivesupp} shows the driving work for an anharmonic oscillator as work medium vs.\ the maximal coupling strength.  With increasing anharmonicity $|W_d|$ decreases and leads to smaller efficiencies (shown in the main text).


\newpage

\newpage

\bibliography{engineMJJ}{}